\documentclass[twocolumn,showpacs,preprintnumbers,prb,aps,superscriptaddress]{revtex4-1}
\usepackage{amsmath}
\usepackage{amssymb}
\usepackage[retain-unity-mantissa=false]{siunitx}
\usepackage{graphicx}
\usepackage{xcolor}
\usepackage[caption=false]{subfig}
\usepackage{placeins}
\usepackage[colorlinks=true, linkcolor=blue, citecolor=blue, filecolor=blue, urlcolor=blue,pdfproducer={.},pdfcreator={.}]{hyperref}
\usepackage{cleveref}
\usepackage{blindtext}
\usepackage{comment}
\usepackage{tikz}
\usepackage{verbatim}
\usepackage{ulem}

\newcommand{\brcom}[1]{} 

\begin{document}

	\title{Phonon-induced long-lasting nonequilibrium in the electron system \\of a laser-excited solid}
	\author{S. T. \surname{Weber}}
	\email{weber@physik.uni-kl.de}
	\affiliation{Department of Physics and Research Center OPTIMAS, Technische Universitaet Kaiserslautern, 67663 Kaiserslautern, Germany}
	\author{B. \surname{Rethfeld}}
	\affiliation{Department of Physics and Research Center OPTIMAS, Technische Universitaet Kaiserslautern, 67663 Kaiserslautern, Germany}
	
	\begin{abstract}
		Electron-electron thermalization and electron-phonon relaxation processes in laser-excited solids are often assumed to occur on different timescales.
		This is true for the majority of the conduction band electrons in a metal. However, electron-phonon interactions can influence the thermalization process of the excited electrons.
		We study the interplay of the underlying scattering mechanisms for the case of a noble metal with help of a set of complete Boltzmann collision integrals. We trace  
		the transient electron distribution in copper and its deviations from a Fermi-Dirac distribution 
		due to the excitation with an ultrashort laser pulse. 
		We investigate the different stages of electronic nonequilibrium after an excitation with an
		ultrashort laser-pulse of \SI{800}{nm} wavelength and \SI{10}{fs} pulse duration. Our calculations show a strong nonequilibrium during and directly after the
		end of the laser pulse. Subsequently, we find a fast thermalization of most electrons. Surprisingly, we observe a long-lasting nonequilibrium, which can be 
		attributed to the electron-phonon scattering. 
		This nonequilibrium establishes at energies around peaks in the density of states of the electrons and
		persists on the timescale of electron-phonon energy relaxation. It influences in turn the electron phonon coupling strength.

	\end{abstract}

	\date{\today}

	\maketitle

	\section{Introduction}

		Ultrashort laser pulses are an essential tool in medical and industrial applications.\cite{Vogel03,Chichkov96,Anisimov2002,BaeuerleBuch11,Vogel2005,Balling2013,Shugaev2016,Rethfeld2017} 
		The development of modern lasers allows for shorter and shorter pulse durations. 
		For material processing, generally applicable descriptions of the energy dissipation in a strongly excited material are needed,
		valid for a large variety of laser parameters and sample sets\cite{Giri2015,Winter2017,Hopkins2008,Lin2008,Carpene2006,Raemer2014}. Such models usually rely on macroscopic parameters like temperatures or densities, in order to provide an easy and
		cost-efficiently prediction of the laser-matter interaction.

		Also fundamental questions of quantum interactions in solids are accessible with ultrashort laser pulses. 
		Intrinsic timescales of excitations are investigated experimentally and theoretically\cite{Bauer2015,Balling2013,Shugaev2016,Aeschlimann1996,Hohlfeld97,DelFatti00,Rameau2016}. 
		First principle calculations are nowadays extended to determine also nonequilibrium states of matter and excited systems
		of quasiparticles, while kinetic methods are applied
		to study the relaxation dynamics of the systems\cite{Dewhurst2018,Recoules2006,Fourment2014,Zijlstra2013,Bernardi2014,Medvedev11,Vorberger2010,Maldonado2017,Baranov2014}.
		Modern high-performance computer clusters open up the possibility to perform
		larger {\it ab initio} calculations tracing also microscopic scattering processes on ultrafast timescales.

		A femtosecond laser pulse drives the electrons out of their thermal equilibrium. Thus, conventional temperature-based models reach their limits to describe the dynamics and the energy dissipation inside the material.
		Kinetic models, like the Boltzmann equation \cite{Rethfeld2002,Sun94,Fann92a,DelFatti00,Knorren2000,Lugovskoy1999} or Monte-Carlo methods
		\cite{Medvedev11,vanHall2001} allow to trace the evolution of the electronic nonequilibrium.
		The electronic thermalization \cite{Fann92a,Sun94,DelFatti00,Lisowski2004,Rethfeld2002,Mueller2013PRB,Silaeva2018} and the influence of the different stages of electronic nonequilibrium 
		distributions on further energy dissipation processes\cite{Bauer2015,Mueller2013PRB,Rethfeld2002,DelFatti00,Baranov2014,Maldonado2017,Weber2017} are in the focus of interest and their details are heavily discussed topics. 
		Recently, also the influence of the nonequilibrium in the phonon system has attracted increasing attention\cite{Waldecker2016,Ono2018,Klett2018}.
		It is common believe, that the relaxation processes proceed in a certain order: First, the laser excites the 
		electrons. Then, the electrons redistribute their energy and thermalize to a Fermi distribution. And finally, the hot electrons heat also the phonon system, i.e. the cold lattice.
		However, the mutual influence between different relaxation mechanisms is rarely addressed.

		In this work, we study simultaneously the electron-electron thermalization and their interaction with the phonons. We confirm the thermalization of most electrons on a 
		timescale of a few tens of femtoseconds. However, we find a long-lasting nonequilibrium in the electron system which is induced by the cold phonons. Surprisingly, this nonequilibrium
		persists on a picosecond timescale. It vanishes, when the electrons and phonons are fully equilibrated with each other.
	
	\section{Model}
	
		We use the time-dependent Boltzmann equation collision integrals
		to study the excitation and thermalization 
		of electrons in a thin copper film, considering
		electron-photon excitation, electron-electron scattering and electron-phonon 
		relaxation.
		Thus, we describe the transient changes of electron distribution $f(E,t)$ 
		and phonon distribution $g(E,t)$\cite{Mueller2013PRB,Weber2017}
		and trace the nonequilibrium effects during and after ultrashort laser excitation. 
		We neglect heat- and particle transport. This case corresponds e.g.~to the investigation 
		of thin copper foils, that are 
		smaller than the depth of homogeneous heating.
		In our approach, we include the excitation of the electrons due to inverse Bremsstrahlung, the thermalization 
		due to electron-electron scattering and the energy transfer from the electronic system to the phononic system.
		Each interaction is described by a full Boltzmann collision integral, which are derived in Ref.~\citenum{Mueller2013PRB}. 
		The corresponding collision terms can be simplified be assuming an isotropic band. 
		Using an effective one-band model \cite{Mueller2013PRB} allows us to implement the realistic density of states (DOS) with reasonable 
		numerical effort. In this model, an averaged isotropic dispersion relation 
		is calculated from the density of states
		taken from density functional theory (DFT)\cite{Lin2008}.
		By that, the influence of distinct features in the DOS can be taken into account \cite{Mueller2013PRB}. 
		The one-band model reflects well the dependence of effective 
		scattering rates on the density of possible states 
		at certain energy levels. It has been applied 
		successfully in magnetic\cite{Mueller2013PRL,Bierbrauer2017} as well as in non-magnetic 
		materials \cite{Mueller2013PRB,Mueller2014,Weber2017}
		to explain effects of nonequilibrium 
		spin-flip scattering processes 
		or electron-phonon energy exchange and their dependence on distinct features 
		of the density of states.		
		
		We evaluate the collision terms as given in Refs.~\citenum{Mueller2013PRB,Weber2017} 
		numerically in each time step and trace the temporal changes of the electron distribution 
		caused by the different considered scattering processes. Our approach allows to 
		study the influence of certain scattering processes separately by neglecting some particular
		collisions.
		Further details on the model can be found in Ref. \citenum{Mueller2013PRB}.
		\begin{table}
			\caption{\label{tab} Material parameters for copper}
		\begin{tabular}{lcccc}
			\hline\hline
			Quantity		&	Symbol		& Unit				& Reference		&	Value	\\\hline
			Speed of sound		&	$c_s$ 		& $\si{\metre\per\second}$	& \onlinecite{CRC05}	&	4760	\\
			Debye energy		&	$E_D$		& meV				& \onlinecite{CRC05}	&	53.8	\\
			Density of states	&	$D(E)$		& \si{eV^{-1}m^{-3}}		& \onlinecite{Lin2008}	&		\\
			\hline\hline
		\end{tabular}
		\end{table}

		Only very few input parameters enter the calculation.
		These are the density of states of the conduction electrons, the volume of the unit cell, 
		and the dispersion relation for the phonons. Here, we apply the Debye model for the 
		phonons\cite{Czycholl}, which is a valid approximation in copper having only acoustic modes.
		The applied material parameters for copper are given in \cref{tab}. 
		 The laser is assumed to have a
		rectangular shaped time profile with a duration of $\tau_L = \SI{10}{\femto\second}$, and a wavelength of $\SI{800}{\nano\metre}$, thus, a photon energy
		of 
		$\hbar\omega = \SI{1.55}{\electronvolt}$. 
		The short rectangular laser pulse allows to distinguish between the laser
		excitation and the relaxation processes of the excited system. 
		The absorbed fluence is \SI{0.65}{mJ/cm^2}. The initial temperature of both, electron and phonons, is at room temperature of $T_0=\SI{300}{K}$.

	\section{Results}
	\subsection{Laser-induced nonequilibrium}

\begin{figure}
				\includegraphics{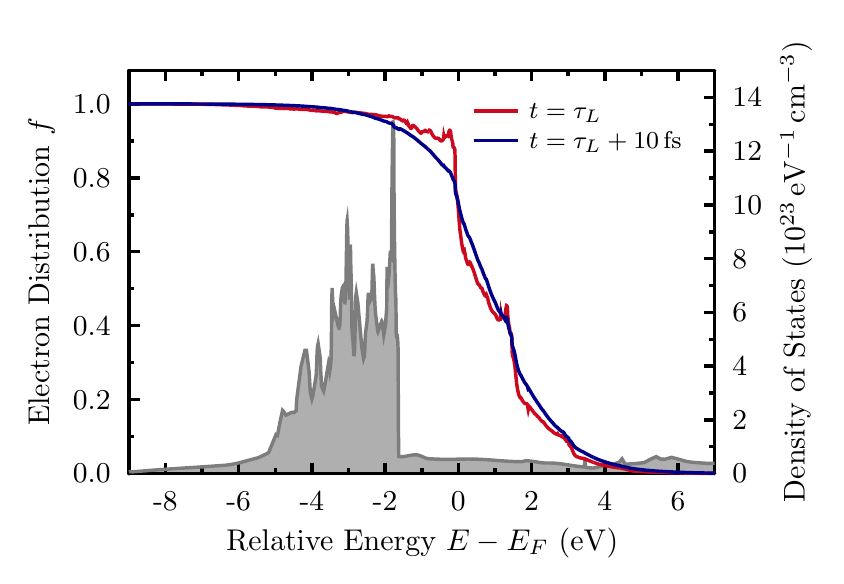}
			\caption{
			Electron distribution of copper at 
			$t=\tau_L$  and at $t=\tau_L + \SI{10}{fs}$. The width of each step corresponds to the photon energy of $\SI{1.55}{eV}$. Distinct features 
			of the density of states are visible in the distribution function. The influence of the electron thermalization
			is already visible. At $t=\tau_L + \SI{10}{fs}$  the distribution almost looks like a Fermi-distribution, but small deviations are still visible.
			\label{fig:transient-f-Cu-10fs}
			}
		\end{figure}

		\Cref{fig:transient-f-Cu-10fs} depicts the electron distribution at the end of the laser pulse ($t=\tau_L=\SI{10}{fs}$) and ten femtoseconds later 
		($t=\tau_L+\SI{10}{fs}$). Directly after the laser pulse, a step-like structure is expected\cite{Rethfeld2002,Knorren2000,Fann92a}. 
		The width of each step corresponds to the photon energy of \SI{1.55}{eV}\cite{Mueller2013PRB,Rethfeld2002}.
		This structure is also reflected here. The steps are not 
		perfectly sharp, due to the thermalization, which takes already place during the very short laser pulse.

		The fast thermalization is a consequence of the high density of excited electrons 
		and the high density of created holes in the {\it d}-bands below the
		Fermi edge, which are induced by the strong excitation fluence. 
		Additionally, DOS-features influence the excited
		distribution function.\cite{Mueller2013PRB,Weber2017} 
	    	Shortly after the end of the laser pulse ($t=\tau_L+\SI{10}{fs}$), the distribution almost looks like a Fermi distribution again. 
		However, nonequilibrium features are still visible. 
		Especially around the high $d$-peak at 
		\SI{-1.8}{eV} below Fermi energy and at the Fermi energy, nonequilibrium features seem to
		persist longer. For a high density of states, small deviations in the distribution can lead to a large number of electrons in nonequilibrium. 

		To quantify these deviations, the total percentage of nonequilibrium electrons,  
		\begin{equation}
			\label{eq:percentage}
			n^{\text{neq}}_{\text{tot}}=\frac{1}{n_e} \int\mathrm{d}E\,  D(E) \lvert f_{\rm neq}(E)-f_{\rm eq}(E)\rvert \enspace,
		\end{equation} 
		is determined, where $n_e$ is the total density of conduction electrons. 
		This value is capable to characterize the strength of electronic nonequilibrium. 
		Its temporal behavior allows for an easy comparison 
		between different stages of thermalization. 
		The corresponding equilibrium distribution $f_{\rm eq}$
		is given by the Fermi-Dirac distribution with the same number of particles and energy content
		as the nonequilibrium distribution $f_{\rm neq}$ at the same instant of time.

		\begin{figure}
			\includegraphics{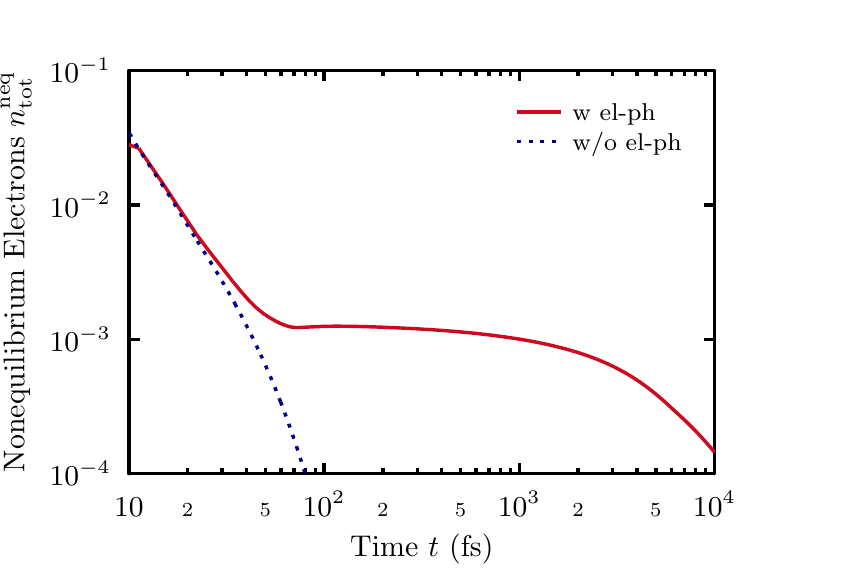}
			\caption{\label{fig:percentage}Percentage of nonequilibrium electrons. Without electron-phonon interactions (blue, dotted line) the thermalization is fast and happens
			on a timescale of tens of femtoseconds. With electron-phonon interactions (red, solid line), the initial thermalization of most electrons is fast. After approximately 
			\SI{60}{fs} a quasi-stationary nonequilibrium establishes. This nonequilibrium decreases within picoseconds due to electron-phonon relaxation.
			}
		\end{figure}

		\begin{figure*}
			\includegraphics{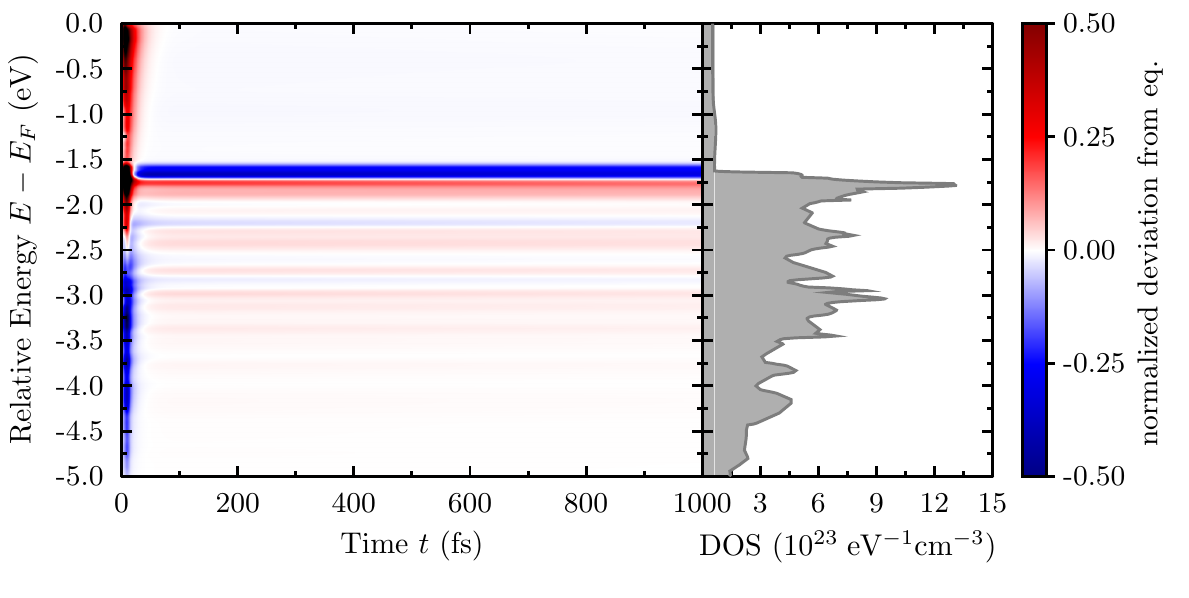}
			\caption{
				Long-lasting nonequilibrium, caused by electron-phonon scattering at large differences in the density
				of states according to \cref{eq:deviation}. Each Peak in the DOS yields to two stripes with the width of a few phonon energies.
				The effect can overlap for peaks with small energetic distance, for example below \SI{-3}{eV}.
			\label{fig:number-diff-Cu-long}}
		\end{figure*}

	\subsection{Phonon-induced nonequilibrium}

		In \cref{fig:percentage} the time evolution of nonequilibrium electrons is shown (red solid line). 
		One can see a decrease of the normalized nonequilibrium electron density corresponding to a
		fast thermalization of most electrons within a few tens of femtoseconds. 
		This agrees with the results of other studies\cite{Mueller2013PRB,Fann92a,Sun94,Knorren2000}
		and the observation in~\cref{fig:transient-f-Cu-10fs}.
		Surprisingly, the electronic nonequilibrium does not vanish completely and the thermalization does not continue further. 
	 	Instead a quasi-stationary nonequilibrium establishes after approximately \SI{60}{fs}. 

		This raises three different questions. First, what causes such a quasi-stationary nonequilibrium? Second, how long does it persist? And third, why does it establish?
		The first two questions can be answered easily. Only electron-electron- and electron-phonon-interactions can have an influence on the electron dynamics after the end of the laser pulse. 
		To test the influence of the phonons, we calculated the dynamics, considering only electron-electron-interactions. 
		The blue, dotted line in \cref{fig:percentage} shows the decrease of the percentage of nonequilibrium
		electrons according to Eq.~(\ref{eq:percentage}). It decreases continuously and thus shows
		the expected behavior. 
		The electron thermalization is fast, with a characteristic
		thermalization time on the range of tens of femtoseconds 
		and no quasi-stationary state appears. 
		Thus, electron-phonon scattering is responsible 
		for the long-lasting nonequilibrium.
		This answers also the second question: As long as the electron-phonon relaxation is not completed, the electronic nonequilibrium persists. \Cref{fig:percentage} shows indeed, 
		that it continues to decrease on a timescale of about ten picoseconds, when electrons and phonons are expected to reach the same temperature.

		The third question concerning the reason behind the quasi-stationary nonequilibrium is more complex to answer. Thus, we have to take a step backwards. The energy-resolved deviation 
		from the equilibrium
		\begin{equation}
			\label{eq:deviation}
			n^{\text{neq}}(E) = \frac{1}{n_e} D(E) \big(f_{\text{neq}}(E)-f_{\text{eq}}(E)\big)
		\end{equation}
		allows to identify the energetic regions in which the quasi-stationary equilibrium persists. 

		\Cref{fig:number-diff-Cu-long} shows this differential nonequilibrium density for timescales 
		during and after laser-excitation on an energy range of \si{5}{eV} below Fermi-edge. 
		The color-coding represents the excess (red) and lack (blue) of electrons, compared to 
		the corresponding equilibrium distribution. 

		Initially, on a timescale of a few tens of femtoseconds, a strong nonequilibrium is visible, 
		cf.~also \cref{fig:transient-f-Cu-10fs}.
		In the energy range directly below Fermi-edge, i.e.~between \SI{0}{eV} and \SI{2.5}{eV},
		 an excess of nonequilibrium electrons can be observed, represented by the red area. 
		Further below Fermi energy, between \SI{2.5}{eV} and \SI{5}{eV}, 
		a lack of electrons exists in the excited nonequilibrium distribution,
		leading to a blue region in \cref{fig:number-diff-Cu-long}. 
		This strong initial nonequilibrium vanishes fast. 

		However, at approximately \SI{60}{fs} a stripe pattern, with alternating areas of 
		lack and excess of electrons appears below Fermi energy. 
		Note that above Fermi-edge no such pattern was observed in our
		calculations. A closer look to the energy range reveals that the largest deviations 
		occur in the energetic area
		of the $d$-band peaks in the density of states.
		These areas of excess and lack of electrons are directly caused by 
		the scattering of electrons with phonons.

		The most prominent stripes are located 
		at a small energy interval 
		around the highest peak in the density of states.
		The small width of the observed stripes
		is a consequence of the small energy of the phonon modes. 
		One phonon can reach energies only
		up to the Debye energy of \SI{53.8}{meV}, for our case of elemental copper. 
		It is clear that in a region of high density of states, more scattering 
		processes are probable and thus the observed resulting feature 
		is more pronounced. 

		However, what is the reason for the alternating deviations from equilibrium? 
		Why are energy regions of excess electrons neighbouring energy regions 
		showing a lack of electrons?
		\begin{figure}
			\includegraphics{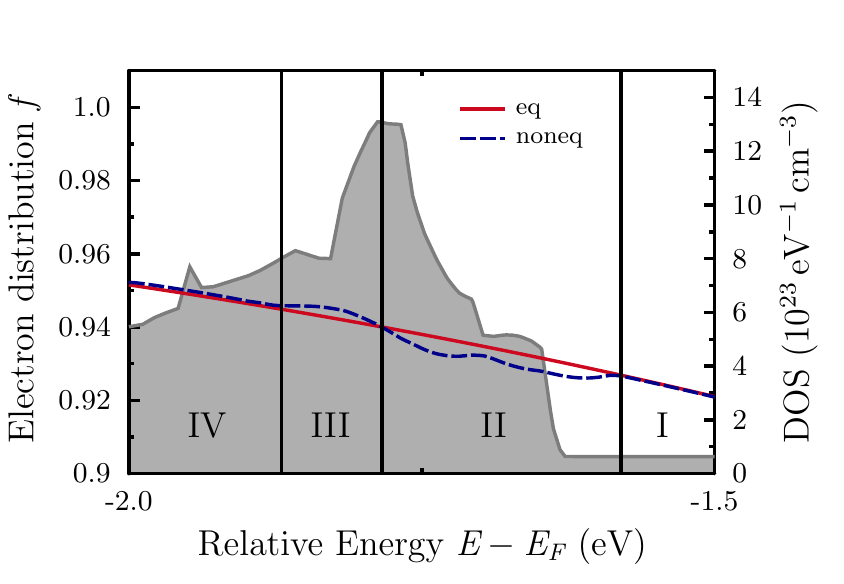}
			\caption{\label{fig:sketch}
			Zoom to the high density of states peak. The area can be divided in four different areas of stronger and weaker electron-phonon interaction.
			Area I and IV have a rather weak, while area II and III have a strong electron-phonon interaction. 
			}
		\end{figure}

		To analyze this further, \cref{fig:sketch} shows a narrow zoom on the distribution function at the highest DOS-peak. 
		The energy axis is divided into four subareas, to explain the mechanisms leading
		to the phonon-induced electronic nonequilibrium. 
		For comparison, \cref{fig:sketch} shows also the corresponding equilibrium electron distribution $f_{\rm eq}$ of the same internal energy.
		Both distributions have a much larger energy content than the phononic system. Thus, electron-phonon scattering will lead to a decrease of the 
		electrons energy. Therefore, we start the discussion with the area of highest kinetic energy proceeding 
		towards lower energies.

		In area I, the density of states is flat, thus the electron-phonon interaction is almost equal over the whole energy range. 
		The (almost) same amount of electrons is scattered in and out 
		of this energy area. That is different in area II. 
		Here, the density of states increases with decreasing energy, 
		which leads to an increased scattering rate. 
		More electrons are scattered to lower-energy areas, than electrons from higher-energy areas 
		can be scattered into this region. 
		A lack of electrons establishes compared to the equilibrium curve. 
		In contrast, for region III, the density of states is decreasing again. Thus, more electrons are scattered into 
		area III from the higher-energy neighboring area II, than
		electrons can be scattered to lower energies.
		An excess of electrons establishes, compared to the equilibrium state. 
		In area IV the slope of the curve changes only slightly, and the in- and out-scattering
		processes are roughly balanced. 

		The long-lasting nonequilibrium is caused by the higher electron-phonon collision rate at the high {\it d}-peaks in the density of states. The small energetic range of the interaction leads to an energetically local
		nonequilibrium, which persist on the timescale of electron-phonon relaxation. 
		We tested the establishing of the long-lasting nonequilibrium with different artificial densities of states. 
		Generally we observe this phenomenon, when oscillations in the 
		density of states occur on energy
		scales in the range of the Debye energy.
		Note that the effect is expected to persist also in case complete band structures
		are considered. Also for momentum-resolved bands, more states are accessible and more
		scattering processes are probable when bands are shallow in the region of 
		electron-phonon interaction.

	\subsection{Influence of the Long-Lasting Nonequilibrium}
	
		In this section, we investigate the influence of the different states of electronic nonequilibrium on the transient electron-phonon coupling parameter 
		$\alpha$. Usually, the electron-phonon coupling parameter is either assumed to be constant, which is a strong simplification and does not fit different
		observations\cite{Lin2008}, or as dependent on the electronic temperature, which is more suitable\cite{Cho2011,Waldecker2016}
		but neglects nonequilibrium effects. With respect to 
		the electronic nonequilibrium, the coupling factor depends on pulse duration, laser wavelength and 
		time.\cite{Mueller2013PRB,Mueller2014,Weber2017} 
		To extract a nonequilibrium coupling parameter, used by temperature descriptions\cite{Anisimov1974,Rethfeld2017},
		we take the temperature $T_e[f_\text{eq}(t)]$ of the corresponding equilibrium distribution with the same
		energy content.\cite{Mueller2013PRB,Weber2017} 
		
		The coupling parameter 
		\begin{equation}
			\alpha = \frac{\text{d} u_e/\text{d} t}{T_p[g(t)]-T_e[f_{eq}(t)]}
		\end{equation}
		can be defined analogously to the two-temperature model.\cite{Anisimov1974}
	
		\begin{figure}
			\includegraphics{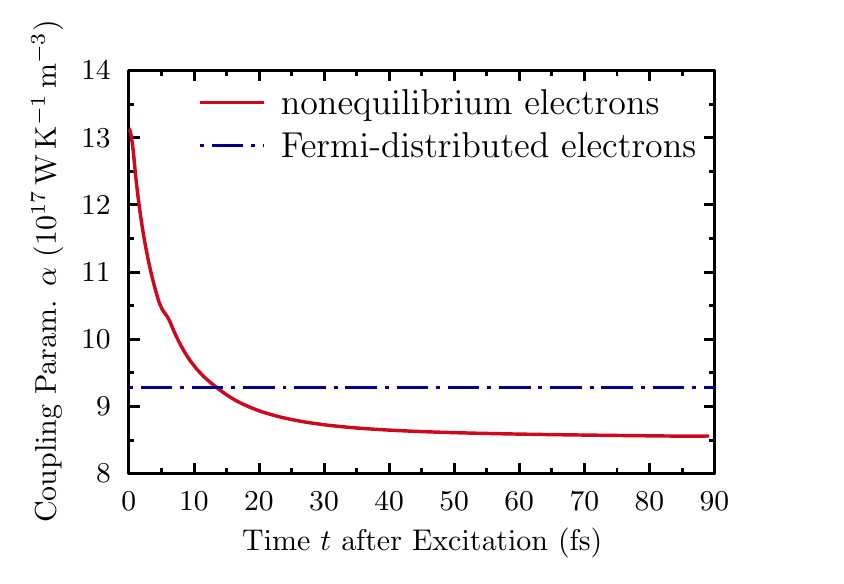}
			\caption{
				Transient electron-phonon coupling parameter $\alpha$. The transient equilibrium parameter decreases slightly within time,
				due to the decreasing electron temperature. At the beginning, the nonequilibrium parameter 
				is higher as the equilibrium parameter and decreases due to electron thermalization. After about ($t=\tau_L+\SI{13}{fs}$) it falls under 
				the equilibrium parameter due to the beginning establishing of the long-lasting nonequilibrium.
			\label{fig:coupling-transient}}
		\end{figure}
	
		The transient coupling parameter is depicted in \Cref{fig:coupling-transient} for equilibrium and nonequilibrium. The equilibrium parameter 
		decreases slightly with time, due to the decreasing electron temperature $T_e$. At first, the nonequilibrium coupling strength exceeds the corresponding equilibrium 
		coupling. Then, due to thermalization within the electronic system, the nonequilibrium decreases and the parameter assimilates to the equilibrium case.  At about 
		($t=\tau_L+\SI{13}{fs}$), the coupling parameters intersect. Without the phonon-induced nonequilibrium, the coupling parameter would match for later times. However, the
		phonons drive the electrons in the long-lasting nonequilibrium state (compare \cref{fig:number-diff-Cu-long,fig:percentage}). This disturbance has also a feedback-effect on the
		electron-phonon coupling on the timescales of energy relaxation. 
		After the parameters intersect, the nonequilibrium parameter falls below the equilibrium coupling, which can be explained with the help of \cref{fig:sketch}.
		In the depicted energy region, the coupling to the phonons in nonequilibrium is weaker as 
		in the equilibrium case as a consequence of the disturbed distribution function. Due to the high density of states and following, the high 
		particle density in this region, it has a high impact on the electron-phonon coupling.

	\section{Summary and Conclusion}

		In this work, we investigated the different timescales and stages of electronic thermalization after ultrashort laser irradiation.
		We applied Boltzmann collision integrals considering
		 the density of states of copper using an effective one-band model.
		We have calculated the energy distribution of the electrons
		excited with an ultrashort laser pulse
		and have traced the thermalization of the laser-induced nonequilibrium.
		The transient electron distribution experiences 
		a fast thermalization in a few tens of femtoseconds. 
		Thus, the electronic subsystem has lost the information of the 
		details of the initial laser excitation on this ultrashort 
		timescale.

		Surprisingly, we have found a long-lasting, quasi-stationary nonequilibrium, 
		which establishes shortly after the end of the laser pulse. 
		This is caused by the phonon emission of hot electrons 
		and persists on the picosecond timescale 
		of electron-phonon relaxation. 
		The nonequilibrium vanishes completely, when electrons and phonons are 
		thermalized.
		A closer look on the energy-resolved deviation of the electron distribution 
		from a Fermi distribution has revealed that an increased electron-phonon scattering 
		at energies around the {\it d}-peaks in the density of states 
		drives the long-lasting nonequilibrium.
		The timescales of equilibration are thus two-fold: a fast initial thermalization
		on femtosecond timescales followed by a quasi-stationary nonequilibrium finally thermalizing 
		on the picosecond timescale. 
		
		We have also shown, that the electron-phonon coupling strength is influenced by the long-lasting
		nonequilibrium in the electronic system. By that, the electronic nonequilibrium can leave a 
		measurable effect on the further energy dissipation.
		
		We thus conclude that even when the timescales of relaxation are
		well separated as in the case of electron-electron thermalization and electron-phonon
		relaxation, the different scattering processes influence each other considerably. 
		As long as the whole system is not fully equilibrated,
                partial equilibria can be disturbed by other interaction partners.

                Financial support of the Deutsche Forschungsgemeinschaft 
                through the Heisenberg grant RE1141-15, the SFB/TRR-173 ``Spin+X" 
		(project A08) 
		and of the Carl-Zeiss Stiftung is gratefully acknowledged.
		We appreciate
                the Allianz f\"{u}r Hochleistungsrechnen Rheinland-Pfalz for providing computing resources through project
                LAINEL on the Elwetritsch high performance computing cluster.
		We thank Uwe Bovensiepen and Nils Brouwer
		for fruitful discussions.

	        \bibliography{draft}

\begin{thebibliography}{48}%
\makeatletter
\providecommand \@ifxundefined [1]{%
 \@ifx{#1\undefined}
}%
\providecommand \@ifnum [1]{%
 \ifnum #1\expandafter \@firstoftwo
 \else \expandafter \@secondoftwo
 \fi
}%
\providecommand \@ifx [1]{%
 \ifx #1\expandafter \@firstoftwo
 \else \expandafter \@secondoftwo
 \fi
}%
\providecommand \natexlab [1]{#1}%
\providecommand \enquote  [1]{``#1''}%
\providecommand \bibnamefont  [1]{#1}%
\providecommand \bibfnamefont [1]{#1}%
\providecommand \citenamefont [1]{#1}%
\providecommand \href@noop [0]{\@secondoftwo}%
\providecommand \href [0]{\begingroup \@sanitize@url \@href}%
\providecommand \@href[1]{\@@startlink{#1}\@@href}%
\providecommand \@@href[1]{\endgroup#1\@@endlink}%
\providecommand \@sanitize@url [0]{\catcode `\\12\catcode `\$12\catcode
  `\&12\catcode `\#12\catcode `\^12\catcode `\_12\catcode `\%12\relax}%
\providecommand \@@startlink[1]{}%
\providecommand \@@endlink[0]{}%
\providecommand \url  [0]{\begingroup\@sanitize@url \@url }%
\providecommand \@url [1]{\endgroup\@href {#1}{\urlprefix }}%
\providecommand \urlprefix  [0]{URL }%
\providecommand \Eprint [0]{\href }%
\providecommand \doibase [0]{http://dx.doi.org/}%
\providecommand \selectlanguage [0]{\@gobble}%
\providecommand \bibinfo  [0]{\@secondoftwo}%
\providecommand \bibfield  [0]{\@secondoftwo}%
\providecommand \translation [1]{[#1]}%
\providecommand \BibitemOpen [0]{}%
\providecommand \bibitemStop [0]{}%
\providecommand \bibitemNoStop [0]{.\EOS\space}%
\providecommand \EOS [0]{\spacefactor3000\relax}%
\providecommand \BibitemShut  [1]{\csname bibitem#1\endcsname}%
\let\auto@bib@innerbib\@empty
\bibitem [{\citenamefont {Vogel}\ and\ \citenamefont
  {Venugopalan}(2003)}]{Vogel03}%
  \BibitemOpen
  \bibfield  {author} {\bibinfo {author} {\bibfnamefont {A.}~\bibnamefont
  {Vogel}}\ and\ \bibinfo {author} {\bibfnamefont {V.}~\bibnamefont
  {Venugopalan}},\ }\href {\doibase 10.1021/cr010379n} {\bibfield  {journal}
  {\bibinfo  {journal} {Chemical Reviews}\ }\textbf {\bibinfo {volume} {103}},\
  \bibinfo {pages} {577} (\bibinfo {year} {2003})}\BibitemShut {NoStop}%
\bibitem [{\citenamefont {Chichkov}\ \emph {et~al.}(1996)\citenamefont
  {Chichkov}, \citenamefont {Momma}, \citenamefont {Nolte}, \citenamefont
  {Alvensleben},\ and\ \citenamefont {T\"{u}nnermann}}]{Chichkov96}%
  \BibitemOpen
  \bibfield  {author} {\bibinfo {author} {\bibfnamefont {B.~N.}\ \bibnamefont
  {Chichkov}}, \bibinfo {author} {\bibfnamefont {C.}~\bibnamefont {Momma}},
  \bibinfo {author} {\bibfnamefont {S.}~\bibnamefont {Nolte}}, \bibinfo
  {author} {\bibfnamefont {F.}~\bibnamefont {Alvensleben}}, \ and\ \bibinfo
  {author} {\bibfnamefont {A.}~\bibnamefont {T\"{u}nnermann}},\ }\href
  {\doibase 10.1007/BF01567637} {\bibfield  {journal} {\bibinfo  {journal}
  {Applied Physics A}\ }\textbf {\bibinfo {volume} {63}},\ \bibinfo {pages}
  {109} (\bibinfo {year} {1996})}\BibitemShut {NoStop}%
\bibitem [{\citenamefont {Anisimov}\ and\ \citenamefont
  {Luk'yanchuk}(2002)}]{Anisimov2002}%
  \BibitemOpen
  \bibfield  {author} {\bibinfo {author} {\bibfnamefont {S.~I.}\ \bibnamefont
  {Anisimov}}\ and\ \bibinfo {author} {\bibfnamefont {B.~S.}\ \bibnamefont
  {Luk'yanchuk}},\ }\href {http://stacks.iop.org/1063-7869/45/i=3/a=R02}
  {\bibfield  {journal} {\bibinfo  {journal} {Physics-Uspekhi}\ }\textbf
  {\bibinfo {volume} {45}},\ \bibinfo {pages} {293} (\bibinfo {year}
  {2002})}\BibitemShut {NoStop}%
\bibitem [{\citenamefont {B\"{a}uerle}(2011)}]{BaeuerleBuch11}%
  \BibitemOpen
  \bibfield  {author} {\bibinfo {author} {\bibfnamefont {D.}~\bibnamefont
  {B\"{a}uerle}},\ }\href@noop {} {\emph {\bibinfo {title} {Laser Processing
  and Chemistry}}}\ (\bibinfo  {publisher} {Springer Verlag},\ \bibinfo
  {address} {Berlin, Heidelberg},\ \bibinfo {year} {2011})\BibitemShut
  {NoStop}%
\bibitem [{\citenamefont {Vogel}\ \emph {et~al.}(2005)\citenamefont {Vogel},
  \citenamefont {Noack}, \citenamefont {H\"{u}ttman},\ and\ \citenamefont
  {Paltauf}}]{Vogel2005}%
  \BibitemOpen
  \bibfield  {author} {\bibinfo {author} {\bibfnamefont {A.}~\bibnamefont
  {Vogel}}, \bibinfo {author} {\bibfnamefont {J.}~\bibnamefont {Noack}},
  \bibinfo {author} {\bibfnamefont {G.}~\bibnamefont {H\"{u}ttman}}, \ and\
  \bibinfo {author} {\bibfnamefont {G.}~\bibnamefont {Paltauf}},\ }\href
  {\doibase https://doi.org/10.1007/s00340-005-2036-6} {\bibfield  {journal}
  {\bibinfo  {journal} {{Appl. Phys. B}}\ }\textbf {\bibinfo {volume} {81}},\
  \bibinfo {pages} {1015} (\bibinfo {year} {2005})}\BibitemShut {NoStop}%
\bibitem [{\citenamefont {Balling}\ and\ \citenamefont
  {Schou}(2013)}]{Balling2013}%
  \BibitemOpen
  \bibfield  {author} {\bibinfo {author} {\bibfnamefont {P.}~\bibnamefont
  {Balling}}\ and\ \bibinfo {author} {\bibfnamefont {J.}~\bibnamefont
  {Schou}},\ }\href {\doibase 10.1088/0034-4885/76/3/036502} {\bibfield
  {journal} {\bibinfo  {journal} {Reports on progress in physics}\ }\textbf
  {\bibinfo {volume} {76}},\ \bibinfo {pages} {036502} (\bibinfo {year}
  {2013})}\BibitemShut {NoStop}%
\bibitem [{\citenamefont {Shugaev}\ \emph {et~al.}(2016)\citenamefont
  {Shugaev}, \citenamefont {Wu}, \citenamefont {Armbruster}, \citenamefont
  {Naghilou}, \citenamefont {Brouwer}, \citenamefont {Ivanov}, \citenamefont
  {Derrien}, \citenamefont {N.M.}, \citenamefont {Kautek}, \citenamefont
  {Rethfeld},\ and\ \citenamefont {Zhigilei}}]{Shugaev2016}%
  \BibitemOpen
  \bibfield  {author} {\bibinfo {author} {\bibfnamefont {M.}~\bibnamefont
  {Shugaev}}, \bibinfo {author} {\bibfnamefont {C.}~\bibnamefont {Wu}},
  \bibinfo {author} {\bibfnamefont {O.}~\bibnamefont {Armbruster}}, \bibinfo
  {author} {\bibfnamefont {A.}~\bibnamefont {Naghilou}}, \bibinfo {author}
  {\bibfnamefont {N.}~\bibnamefont {Brouwer}}, \bibinfo {author} {\bibfnamefont
  {D.}~\bibnamefont {Ivanov}}, \bibinfo {author} {\bibfnamefont {T.~J.-Y.}\
  \bibnamefont {Derrien}}, \bibinfo {author} {\bibfnamefont {B.}~\bibnamefont
  {N.M.}}, \bibinfo {author} {\bibfnamefont {W.}~\bibnamefont {Kautek}},
  \bibinfo {author} {\bibfnamefont {B.}~\bibnamefont {Rethfeld}}, \ and\
  \bibinfo {author} {\bibfnamefont {L.}~\bibnamefont {Zhigilei}},\ }\href
  {\doibase 10.1557/mrs.2016.274} {\bibfield  {journal} {\bibinfo  {journal}
  {MRS Bulletin}\ }\textbf {\bibinfo {volume} {41}},\ \bibinfo {pages} {960}
  (\bibinfo {year} {2016})}\BibitemShut {NoStop}%
\bibitem [{\citenamefont {Rethfeld}\ \emph {et~al.}(2017)\citenamefont
  {Rethfeld}, \citenamefont {Ivanov}, \citenamefont {Garcia},\ and\
  \citenamefont {Anisimov}}]{Rethfeld2017}%
  \BibitemOpen
  \bibfield  {author} {\bibinfo {author} {\bibfnamefont {B.}~\bibnamefont
  {Rethfeld}}, \bibinfo {author} {\bibfnamefont {D.~S.}\ \bibnamefont
  {Ivanov}}, \bibinfo {author} {\bibfnamefont {M.~E.}\ \bibnamefont {Garcia}},
  \ and\ \bibinfo {author} {\bibfnamefont {S.~I.}\ \bibnamefont {Anisimov}},\
  }\href {\doibase 10.1088/1361-6463/50/19/193001} {\bibfield  {journal}
  {\bibinfo  {journal} {Journal of Physics D: Applied Physics}\ }\textbf
  {\bibinfo {volume} {50}},\ \bibinfo {pages} {193001} (\bibinfo {year}
  {2017})}\BibitemShut {NoStop}%
\bibitem [{\citenamefont {Giri}\ \emph {et~al.}(2015)\citenamefont {Giri},
  \citenamefont {Gaskins}, \citenamefont {Foley}, \citenamefont {Cheaito},\
  and\ \citenamefont {Hopkins}}]{Giri2015}%
  \BibitemOpen
  \bibfield  {author} {\bibinfo {author} {\bibfnamefont {A.}~\bibnamefont
  {Giri}}, \bibinfo {author} {\bibfnamefont {J.~T.}\ \bibnamefont {Gaskins}},
  \bibinfo {author} {\bibfnamefont {B.~M.}\ \bibnamefont {Foley}}, \bibinfo
  {author} {\bibfnamefont {R.}~\bibnamefont {Cheaito}}, \ and\ \bibinfo
  {author} {\bibfnamefont {P.~E.}\ \bibnamefont {Hopkins}},\ }\href {\doibase
  http://dx.doi.org/10.1063/1.4906553} {\bibfield  {journal} {\bibinfo
  {journal} {Journal of Applied Physics}\ }\textbf {\bibinfo {volume} {117}},\
  \bibinfo {pages} {044305} (\bibinfo {year} {2015})}\BibitemShut {NoStop}%
\bibitem [{\citenamefont {Winter}(2017)}]{Winter2017}%
  \BibitemOpen
  \bibfield  {author} {\bibinfo {author} {\bibfnamefont {J.}~\bibnamefont
  {Winter}},\ }\href {\doibase 10.1016/j.apsusc.2017.02.070} {\bibfield
  {journal} {\bibinfo  {journal} {Applied Surface Science}\ }\textbf {\bibinfo
  {volume} {417}},\ \bibinfo {pages} {2} (\bibinfo {year} {2017})}\BibitemShut
  {NoStop}%
\bibitem [{\citenamefont {Hopkins}(2008)}]{Hopkins2008}%
  \BibitemOpen
  \bibfield  {author} {\bibinfo {author} {\bibfnamefont {P.~E.}\ \bibnamefont
  {Hopkins}},\ }\href {\doibase 10.1080/15567260802591985} {\bibfield
  {journal} {\bibinfo  {journal} {NMTE}\ }\textbf {\bibinfo {volume} {12}},\
  \bibinfo {pages} {320} (\bibinfo {year} {2008})}\BibitemShut {NoStop}%
\bibitem [{\citenamefont {Lin}\ \emph {et~al.}(2008)\citenamefont {Lin},
  \citenamefont {Zhigilei},\ and\ \citenamefont {Celli}}]{Lin2008}%
  \BibitemOpen
  \bibfield  {author} {\bibinfo {author} {\bibfnamefont {Z.}~\bibnamefont
  {Lin}}, \bibinfo {author} {\bibfnamefont {L.~V.}\ \bibnamefont {Zhigilei}}, \
  and\ \bibinfo {author} {\bibfnamefont {V.}~\bibnamefont {Celli}},\ }\href
  {\doibase 10.1103/PhysRevB.77.075133} {\bibfield  {journal} {\bibinfo
  {journal} {Phys. Rev. B}\ }\textbf {\bibinfo {volume} {77}},\ \bibinfo
  {pages} {075133} (\bibinfo {year} {2008})}\BibitemShut {NoStop}%
\bibitem [{\citenamefont {Carpene}(2006)}]{Carpene2006}%
  \BibitemOpen
  \bibfield  {author} {\bibinfo {author} {\bibfnamefont {E.}~\bibnamefont
  {Carpene}},\ }\href {\doibase 10.1103/PhysRevB.74.024301} {\bibfield
  {journal} {\bibinfo  {journal} {Phys. Rev. B}\ }\textbf {\bibinfo {volume}
  {74}},\ \bibinfo {pages} {024301} (\bibinfo {year} {2006})}\BibitemShut
  {NoStop}%
\bibitem [{\citenamefont {R\"{a}mer}\ \emph {et~al.}(2014)\citenamefont
  {R\"{a}mer}, \citenamefont {Osmani},\ and\ \citenamefont
  {Rethfeld}}]{Raemer2014}%
  \BibitemOpen
  \bibfield  {author} {\bibinfo {author} {\bibfnamefont {A.}~\bibnamefont
  {R\"{a}mer}}, \bibinfo {author} {\bibfnamefont {O.}~\bibnamefont {Osmani}}, \
  and\ \bibinfo {author} {\bibfnamefont {B.}~\bibnamefont {Rethfeld}},\ }\href
  {\doibase 10.1063/1.4891633} {\bibfield  {journal} {\bibinfo  {journal}
  {Journal of Applied Physics}\ }\textbf {\bibinfo {volume} {116}},\ \bibinfo
  {pages} {053508} (\bibinfo {year} {2014})}\BibitemShut {NoStop}%
\bibitem [{\citenamefont {Bauer}\ \emph {et~al.}(2015)\citenamefont {Bauer},
  \citenamefont {Marienfeld},\ and\ \citenamefont {Aeschlimann}}]{Bauer2015}%
  \BibitemOpen
  \bibfield  {author} {\bibinfo {author} {\bibfnamefont {M.}~\bibnamefont
  {Bauer}}, \bibinfo {author} {\bibfnamefont {A.}~\bibnamefont {Marienfeld}}, \
  and\ \bibinfo {author} {\bibfnamefont {M.}~\bibnamefont {Aeschlimann}},\
  }\href {\doibase https://doi.org/10.1016/j.progsurf.2015.05.001} {\bibfield
  {journal} {\bibinfo  {journal} {Progress in Surface Science}\ }\textbf
  {\bibinfo {volume} {90}},\ \bibinfo {pages} {319 } (\bibinfo {year}
  {2015})}\BibitemShut {NoStop}%
\bibitem [{\citenamefont {Aeschlimann}\ \emph {et~al.}(1996)\citenamefont
  {Aeschlimann}, \citenamefont {Bauer},\ and\ \citenamefont
  {Pawlik}}]{Aeschlimann1996}%
  \BibitemOpen
  \bibfield  {author} {\bibinfo {author} {\bibfnamefont {M.}~\bibnamefont
  {Aeschlimann}}, \bibinfo {author} {\bibfnamefont {M.}~\bibnamefont {Bauer}},
  \ and\ \bibinfo {author} {\bibfnamefont {S.}~\bibnamefont {Pawlik}},\ }\href
  {http://www.ieap.uni-kiel.de/solid/ag-bauer/pdf/paper_00000001.pdf}
  {\bibfield  {journal} {\bibinfo  {journal} {Chemical Physics}\ }\textbf
  {\bibinfo {volume} {205}},\ \bibinfo {pages} {127} (\bibinfo {year}
  {1996})}\BibitemShut {NoStop}%
\bibitem [{\citenamefont {Hohlfeld}\ \emph {et~al.}(1997)\citenamefont
  {Hohlfeld}, \citenamefont {M\"uller}, \citenamefont {Wellershoff},\ and\
  \citenamefont {Matthias}}]{Hohlfeld97}%
  \BibitemOpen
  \bibfield  {author} {\bibinfo {author} {\bibfnamefont {J.}~\bibnamefont
  {Hohlfeld}}, \bibinfo {author} {\bibfnamefont {J.~G.}\ \bibnamefont
  {M\"uller}}, \bibinfo {author} {\bibfnamefont {S.-S.}\ \bibnamefont
  {Wellershoff}}, \ and\ \bibinfo {author} {\bibfnamefont {E.}~\bibnamefont
  {Matthias}},\ }\href {\doibase 10.1007/s003400050189} {\bibfield  {journal}
  {\bibinfo  {journal} {Appl. Phys. B}\ }\textbf {\bibinfo {volume} {64}},\
  \bibinfo {pages} {387} (\bibinfo {year} {1997})}\BibitemShut {NoStop}%
\bibitem [{\citenamefont {Del~Fatti}\ \emph {et~al.}(2000)\citenamefont
  {Del~Fatti}, \citenamefont {Voisin}, \citenamefont {Achermann}, \citenamefont
  {Tzortzakis}, \citenamefont {Christofilos},\ and\ \citenamefont
  {Vall\'ee}}]{DelFatti00}%
  \BibitemOpen
  \bibfield  {author} {\bibinfo {author} {\bibfnamefont {N.}~\bibnamefont
  {Del~Fatti}}, \bibinfo {author} {\bibfnamefont {C.}~\bibnamefont {Voisin}},
  \bibinfo {author} {\bibfnamefont {M.}~\bibnamefont {Achermann}}, \bibinfo
  {author} {\bibfnamefont {S.}~\bibnamefont {Tzortzakis}}, \bibinfo {author}
  {\bibfnamefont {D.}~\bibnamefont {Christofilos}}, \ and\ \bibinfo {author}
  {\bibfnamefont {F.}~\bibnamefont {Vall\'ee}},\ }\href {\doibase
  10.1103/PhysRevB.61.16956} {\bibfield  {journal} {\bibinfo  {journal} {Phys.
  Rev. B}\ }\textbf {\bibinfo {volume} {61}},\ \bibinfo {pages} {16956}
  (\bibinfo {year} {2000})}\BibitemShut {NoStop}%
\bibitem [{\citenamefont {Rameau}\ \emph {et~al.}(2016)\citenamefont {Rameau},
  \citenamefont {Freutel}, \citenamefont {Kemper}, \citenamefont {Sentef},
  \citenamefont {Freericks}, \citenamefont {Avigo}, \citenamefont {Ligges},
  \citenamefont {Rettig}, \citenamefont {Yoshida}, \citenamefont {Eisaki} \emph
  {et~al.}}]{Rameau2016}%
  \BibitemOpen
  \bibfield  {author} {\bibinfo {author} {\bibfnamefont {J.}~\bibnamefont
  {Rameau}}, \bibinfo {author} {\bibfnamefont {S.}~\bibnamefont {Freutel}},
  \bibinfo {author} {\bibfnamefont {A.}~\bibnamefont {Kemper}}, \bibinfo
  {author} {\bibfnamefont {M.}~\bibnamefont {Sentef}}, \bibinfo {author}
  {\bibfnamefont {J.}~\bibnamefont {Freericks}}, \bibinfo {author}
  {\bibfnamefont {I.}~\bibnamefont {Avigo}}, \bibinfo {author} {\bibfnamefont
  {M.}~\bibnamefont {Ligges}}, \bibinfo {author} {\bibfnamefont
  {L.}~\bibnamefont {Rettig}}, \bibinfo {author} {\bibfnamefont
  {Y.}~\bibnamefont {Yoshida}}, \bibinfo {author} {\bibfnamefont
  {H.}~\bibnamefont {Eisaki}},  \emph {et~al.},\ }\href {\doibase
  10.1038/ncomms13761} {\bibfield  {journal} {\bibinfo  {journal} {Nature
  communications}\ }\textbf {\bibinfo {volume} {7}},\ \bibinfo {pages} {13761}
  (\bibinfo {year} {2016})}\BibitemShut {NoStop}%
\bibitem [{\citenamefont {Dewhurst}\ \emph {et~al.}(2018)\citenamefont
  {Dewhurst}, \citenamefont {Elliott}, \citenamefont {Shallcross},
  \citenamefont {Gross},\ and\ \citenamefont {Sharma}}]{Dewhurst2018}%
  \BibitemOpen
  \bibfield  {author} {\bibinfo {author} {\bibfnamefont {J.~K.}\ \bibnamefont
  {Dewhurst}}, \bibinfo {author} {\bibfnamefont {P.}~\bibnamefont {Elliott}},
  \bibinfo {author} {\bibfnamefont {S.}~\bibnamefont {Shallcross}}, \bibinfo
  {author} {\bibfnamefont {E.~K.}\ \bibnamefont {Gross}}, \ and\ \bibinfo
  {author} {\bibfnamefont {S.}~\bibnamefont {Sharma}},\ }\href {\doibase
  10.1021/acs.nanolett.7b05118} {\bibfield  {journal} {\bibinfo  {journal}
  {Nano letters}\ }\textbf {\bibinfo {volume} {18}},\ \bibinfo {pages} {1842}
  (\bibinfo {year} {2018})}\BibitemShut {NoStop}%
\bibitem [{\citenamefont {Recoules}\ \emph {et~al.}(2006)\citenamefont
  {Recoules}, \citenamefont {Cl\'erouin}, \citenamefont {Z\'erah},
  \citenamefont {Anglade},\ and\ \citenamefont {Mazevet}}]{Recoules2006}%
  \BibitemOpen
  \bibfield  {author} {\bibinfo {author} {\bibfnamefont {V.}~\bibnamefont
  {Recoules}}, \bibinfo {author} {\bibfnamefont {J.}~\bibnamefont
  {Cl\'erouin}}, \bibinfo {author} {\bibfnamefont {G.}~\bibnamefont {Z\'erah}},
  \bibinfo {author} {\bibfnamefont {P.~M.}\ \bibnamefont {Anglade}}, \ and\
  \bibinfo {author} {\bibfnamefont {S.}~\bibnamefont {Mazevet}},\ }\href
  {\doibase 10.1103/PhysRevLett.96.055503} {\bibfield  {journal} {\bibinfo
  {journal} {Phys. Rev. Lett.}\ }\textbf {\bibinfo {volume} {96}},\ \bibinfo
  {pages} {055503} (\bibinfo {year} {2006})}\BibitemShut {NoStop}%
\bibitem [{\citenamefont {Fourment}\ \emph {et~al.}(2014)\citenamefont
  {Fourment}, \citenamefont {Deneuville}, \citenamefont {Descamps},
  \citenamefont {Dorchies}, \citenamefont {Petit}, \citenamefont {Peyrusse},
  \citenamefont {Holst},\ and\ \citenamefont {Recoules}}]{Fourment2014}%
  \BibitemOpen
  \bibfield  {author} {\bibinfo {author} {\bibfnamefont {C.}~\bibnamefont
  {Fourment}}, \bibinfo {author} {\bibfnamefont {F.}~\bibnamefont
  {Deneuville}}, \bibinfo {author} {\bibfnamefont {D.}~\bibnamefont
  {Descamps}}, \bibinfo {author} {\bibfnamefont {F.}~\bibnamefont {Dorchies}},
  \bibinfo {author} {\bibfnamefont {S.}~\bibnamefont {Petit}}, \bibinfo
  {author} {\bibfnamefont {O.}~\bibnamefont {Peyrusse}}, \bibinfo {author}
  {\bibfnamefont {B.}~\bibnamefont {Holst}}, \ and\ \bibinfo {author}
  {\bibfnamefont {V.}~\bibnamefont {Recoules}},\ }\href {\doibase
  10.1103/PhysRevB.89.161110} {\bibfield  {journal} {\bibinfo  {journal} {Phys.
  Rev. B}\ }\textbf {\bibinfo {volume} {89}},\ \bibinfo {pages} {161110}
  (\bibinfo {year} {2014})}\BibitemShut {NoStop}%
\bibitem [{\citenamefont {Zijlstra}\ \emph {et~al.}(2013)\citenamefont
  {Zijlstra}, \citenamefont {Cheenicode~Kabeer}, \citenamefont {Bauerhenne},
  \citenamefont {Zier}, \citenamefont {Grigoryan},\ and\ \citenamefont
  {Garcia}}]{Zijlstra2013}%
  \BibitemOpen
  \bibfield  {author} {\bibinfo {author} {\bibfnamefont {E.~S.}\ \bibnamefont
  {Zijlstra}}, \bibinfo {author} {\bibfnamefont {F.}~\bibnamefont
  {Cheenicode~Kabeer}}, \bibinfo {author} {\bibfnamefont {B.}~\bibnamefont
  {Bauerhenne}}, \bibinfo {author} {\bibfnamefont {T.}~\bibnamefont {Zier}},
  \bibinfo {author} {\bibfnamefont {N.}~\bibnamefont {Grigoryan}}, \ and\
  \bibinfo {author} {\bibfnamefont {M.~E.}\ \bibnamefont {Garcia}},\ }\href
  {\doibase 10.1007/s00339-012-7183-0} {\bibfield  {journal} {\bibinfo
  {journal} {Applied Physics A}\ }\textbf {\bibinfo {volume} {110}},\ \bibinfo
  {pages} {519} (\bibinfo {year} {2013})}\BibitemShut {NoStop}%
\bibitem [{\citenamefont {Bernardi}\ \emph {et~al.}(2014)\citenamefont
  {Bernardi}, \citenamefont {Vigil-Fowler}, \citenamefont {Lischner},
  \citenamefont {Neaton},\ and\ \citenamefont {Louie}}]{Bernardi2014}%
  \BibitemOpen
  \bibfield  {author} {\bibinfo {author} {\bibfnamefont {M.}~\bibnamefont
  {Bernardi}}, \bibinfo {author} {\bibfnamefont {D.}~\bibnamefont
  {Vigil-Fowler}}, \bibinfo {author} {\bibfnamefont {J.}~\bibnamefont
  {Lischner}}, \bibinfo {author} {\bibfnamefont {J.~B.}\ \bibnamefont
  {Neaton}}, \ and\ \bibinfo {author} {\bibfnamefont {S.~G.}\ \bibnamefont
  {Louie}},\ }\href {\doibase 10.1103/PhysRevLett.112.257402} {\bibfield
  {journal} {\bibinfo  {journal} {Phys. Rev. Lett.}\ }\textbf {\bibinfo
  {volume} {112}},\ \bibinfo {pages} {257402} (\bibinfo {year}
  {2014})}\BibitemShut {NoStop}%
\bibitem [{\citenamefont {Medvedev}\ \emph {et~al.}(2011)\citenamefont
  {Medvedev}, \citenamefont {Zastrau}, \citenamefont {F\"orster}, \citenamefont
  {Gericke},\ and\ \citenamefont {Rethfeld}}]{Medvedev11}%
  \BibitemOpen
  \bibfield  {author} {\bibinfo {author} {\bibfnamefont {N.}~\bibnamefont
  {Medvedev}}, \bibinfo {author} {\bibfnamefont {U.}~\bibnamefont {Zastrau}},
  \bibinfo {author} {\bibfnamefont {E.}~\bibnamefont {F\"orster}}, \bibinfo
  {author} {\bibfnamefont {D.~O.}\ \bibnamefont {Gericke}}, \ and\ \bibinfo
  {author} {\bibfnamefont {B.}~\bibnamefont {Rethfeld}},\ }\href {\doibase
  10.1103/PhysRevLett.107.165003} {\bibfield  {journal} {\bibinfo  {journal}
  {Phys. Rev. Lett.}\ }\textbf {\bibinfo {volume} {107}},\ \bibinfo {pages}
  {165003} (\bibinfo {year} {2011})}\BibitemShut {NoStop}%
\bibitem [{\citenamefont {Vorberger}\ \emph {et~al.}(2010)\citenamefont
  {Vorberger}, \citenamefont {Gericke}, \citenamefont {Bornath},\ and\
  \citenamefont {Schlanges}}]{Vorberger2010}%
  \BibitemOpen
  \bibfield  {author} {\bibinfo {author} {\bibfnamefont {J.}~\bibnamefont
  {Vorberger}}, \bibinfo {author} {\bibfnamefont {D.~O.}\ \bibnamefont
  {Gericke}}, \bibinfo {author} {\bibfnamefont {T.}~\bibnamefont {Bornath}}, \
  and\ \bibinfo {author} {\bibfnamefont {M.}~\bibnamefont {Schlanges}},\ }\href
  {\doibase 10.1103/PhysRevE.81.046404} {\bibfield  {journal} {\bibinfo
  {journal} {Phys. Rev. E}\ }\textbf {\bibinfo {volume} {81}},\ \bibinfo
  {pages} {046404} (\bibinfo {year} {2010})}\BibitemShut {NoStop}%
\bibitem [{\citenamefont {Maldonado}\ \emph {et~al.}(2017)\citenamefont
  {Maldonado}, \citenamefont {Carva}, \citenamefont {Flammer},\ and\
  \citenamefont {Oppeneer}}]{Maldonado2017}%
  \BibitemOpen
  \bibfield  {author} {\bibinfo {author} {\bibfnamefont {P.}~\bibnamefont
  {Maldonado}}, \bibinfo {author} {\bibfnamefont {K.}~\bibnamefont {Carva}},
  \bibinfo {author} {\bibfnamefont {M.}~\bibnamefont {Flammer}}, \ and\
  \bibinfo {author} {\bibfnamefont {P.~M.}\ \bibnamefont {Oppeneer}},\ }\href
  {\doibase 10.1103/PhysRevB.96.174439} {\bibfield  {journal} {\bibinfo
  {journal} {Phys. Rev. B}\ }\textbf {\bibinfo {volume} {96}},\ \bibinfo
  {pages} {174439} (\bibinfo {year} {2017})}\BibitemShut {NoStop}%
\bibitem [{\citenamefont {Baranov}\ and\ \citenamefont
  {Kabanov}(2014)}]{Baranov2014}%
  \BibitemOpen
  \bibfield  {author} {\bibinfo {author} {\bibfnamefont {V.~V.}\ \bibnamefont
  {Baranov}}\ and\ \bibinfo {author} {\bibfnamefont {V.~V.}\ \bibnamefont
  {Kabanov}},\ }\href {\doibase 10.1103/PhysRevB.89.125102} {\bibfield
  {journal} {\bibinfo  {journal} {Physical Review B}\ }\textbf {\bibinfo
  {volume} {89}},\ \bibinfo {pages} {125102} (\bibinfo {year}
  {2014})}\BibitemShut {NoStop}%
\bibitem [{\citenamefont {Rethfeld}\ \emph {et~al.}(2002)\citenamefont
  {Rethfeld}, \citenamefont {Kaiser}, \citenamefont {Vicanek},\ and\
  \citenamefont {Simon}}]{Rethfeld2002}%
  \BibitemOpen
  \bibfield  {author} {\bibinfo {author} {\bibfnamefont {B.}~\bibnamefont
  {Rethfeld}}, \bibinfo {author} {\bibfnamefont {A.}~\bibnamefont {Kaiser}},
  \bibinfo {author} {\bibfnamefont {M.}~\bibnamefont {Vicanek}}, \ and\
  \bibinfo {author} {\bibfnamefont {G.}~\bibnamefont {Simon}},\ }\href
  {\doibase 10.1103/PhysRevB.65.214303} {\bibfield  {journal} {\bibinfo
  {journal} {Phys. Rev. B}\ }\textbf {\bibinfo {volume} {65}},\ \bibinfo
  {pages} {214303} (\bibinfo {year} {2002})}\BibitemShut {NoStop}%
\bibitem [{\citenamefont {Sun}\ \emph {et~al.}(1994)\citenamefont {Sun},
  \citenamefont {Vall\'ee}, \citenamefont {Acioli}, \citenamefont {Ippen},\
  and\ \citenamefont {Fujimoto}}]{Sun94}%
  \BibitemOpen
  \bibfield  {author} {\bibinfo {author} {\bibfnamefont {C.-K.}\ \bibnamefont
  {Sun}}, \bibinfo {author} {\bibfnamefont {F.}~\bibnamefont {Vall\'ee}},
  \bibinfo {author} {\bibfnamefont {L.~H.}\ \bibnamefont {Acioli}}, \bibinfo
  {author} {\bibfnamefont {E.~P.}\ \bibnamefont {Ippen}}, \ and\ \bibinfo
  {author} {\bibfnamefont {J.~G.}\ \bibnamefont {Fujimoto}},\ }\href {\doibase
  10.1103/PhysRevB.50.15337} {\bibfield  {journal} {\bibinfo  {journal} {Phys.
  Rev. B}\ }\textbf {\bibinfo {volume} {50}},\ \bibinfo {pages} {15337}
  (\bibinfo {year} {1994})}\BibitemShut {NoStop}%
\bibitem [{\citenamefont {Fann}\ \emph {et~al.}(1992)\citenamefont {Fann},
  \citenamefont {Storz}, \citenamefont {Tom},\ and\ \citenamefont
  {Bokor}}]{Fann92a}%
  \BibitemOpen
  \bibfield  {author} {\bibinfo {author} {\bibfnamefont {W.~S.}\ \bibnamefont
  {Fann}}, \bibinfo {author} {\bibfnamefont {R.}~\bibnamefont {Storz}},
  \bibinfo {author} {\bibfnamefont {H.~W.~K.}\ \bibnamefont {Tom}}, \ and\
  \bibinfo {author} {\bibfnamefont {J.}~\bibnamefont {Bokor}},\ }\href
  {\doibase 10.1103/PhysRevB.46.13592} {\bibfield  {journal} {\bibinfo
  {journal} {Phys. Rev. B}\ }\textbf {\bibinfo {volume} {46}},\ \bibinfo
  {pages} {13592} (\bibinfo {year} {1992})}\BibitemShut {NoStop}%
\bibitem [{\citenamefont {Knorren}\ \emph {et~al.}(2000)\citenamefont
  {Knorren}, \citenamefont {Bennemann}, \citenamefont {Burgermeister},\ and\
  \citenamefont {Aeschlimann}}]{Knorren2000}%
  \BibitemOpen
  \bibfield  {author} {\bibinfo {author} {\bibfnamefont {R.}~\bibnamefont
  {Knorren}}, \bibinfo {author} {\bibfnamefont {K.~H.}\ \bibnamefont
  {Bennemann}}, \bibinfo {author} {\bibfnamefont {R.}~\bibnamefont
  {Burgermeister}}, \ and\ \bibinfo {author} {\bibfnamefont {M.}~\bibnamefont
  {Aeschlimann}},\ }\href {\doibase 10.1103/PhysRevB.61.9427} {\bibfield
  {journal} {\bibinfo  {journal} {Phys. Rev. B}\ }\textbf {\bibinfo {volume}
  {61}},\ \bibinfo {pages} {9427} (\bibinfo {year} {2000})}\BibitemShut
  {NoStop}%
\bibitem [{\citenamefont {Lugovskoy}\ and\ \citenamefont
  {Bray}(1999)}]{Lugovskoy1999}%
  \BibitemOpen
  \bibfield  {author} {\bibinfo {author} {\bibfnamefont {A.~V.}\ \bibnamefont
  {Lugovskoy}}\ and\ \bibinfo {author} {\bibfnamefont {I.}~\bibnamefont
  {Bray}},\ }\href {\doibase 10.1103/PhysRevB.60.3279} {\bibfield  {journal}
  {\bibinfo  {journal} {Phys. Rev. B}\ }\textbf {\bibinfo {volume} {60}},\
  \bibinfo {pages} {3279} (\bibinfo {year} {1999})}\BibitemShut {NoStop}%
\bibitem [{\citenamefont {van Hall}(2001)}]{vanHall2001}%
  \BibitemOpen
  \bibfield  {author} {\bibinfo {author} {\bibfnamefont {P.~J.}\ \bibnamefont
  {van Hall}},\ }\href {\doibase 10.1103/PhysRevB.63.104301} {\bibfield
  {journal} {\bibinfo  {journal} {Phys. Rev. B}\ }\textbf {\bibinfo {volume}
  {63}},\ \bibinfo {pages} {104301} (\bibinfo {year} {2001})}\BibitemShut
  {NoStop}%
\bibitem [{\citenamefont {Lisowski}\ \emph {et~al.}(2004)\citenamefont
  {Lisowski}, \citenamefont {Loukakos}, \citenamefont {Bovensiepen},
  \citenamefont {St{\"a}hler}, \citenamefont {Gahl},\ and\ \citenamefont
  {Wolf}}]{Lisowski2004}%
  \BibitemOpen
  \bibfield  {author} {\bibinfo {author} {\bibfnamefont {M.}~\bibnamefont
  {Lisowski}}, \bibinfo {author} {\bibfnamefont {P.}~\bibnamefont {Loukakos}},
  \bibinfo {author} {\bibfnamefont {U.}~\bibnamefont {Bovensiepen}}, \bibinfo
  {author} {\bibfnamefont {J.}~\bibnamefont {St{\"a}hler}}, \bibinfo {author}
  {\bibfnamefont {C.}~\bibnamefont {Gahl}}, \ and\ \bibinfo {author}
  {\bibfnamefont {M.}~\bibnamefont {Wolf}},\ }\href {\doibase
  10.1007/s00339-003-2301-7} {\bibfield  {journal} {\bibinfo  {journal}
  {Applied Physics A}\ }\textbf {\bibinfo {volume} {78}},\ \bibinfo {pages}
  {165} (\bibinfo {year} {2004})}\BibitemShut {NoStop}%
\bibitem [{\citenamefont {Mueller}\ and\ \citenamefont
  {Rethfeld}(2013)}]{Mueller2013PRB}%
  \BibitemOpen
  \bibfield  {author} {\bibinfo {author} {\bibfnamefont {B.~Y.}\ \bibnamefont
  {Mueller}}\ and\ \bibinfo {author} {\bibfnamefont {B.}~\bibnamefont
  {Rethfeld}},\ }\href {\doibase 10.1103/PhysRevB.87.035139} {\bibfield
  {journal} {\bibinfo  {journal} {Phys. Rev. B}\ }\textbf {\bibinfo {volume}
  {87}},\ \bibinfo {pages} {035139} (\bibinfo {year} {2013})}\BibitemShut
  {NoStop}%
\bibitem [{\citenamefont {Silaeva}\ \emph {et~al.}(2018)\citenamefont
  {Silaeva}, \citenamefont {Bevillon}, \citenamefont {Stoian},\ and\
  \citenamefont {Colombier}}]{Silaeva2018}%
  \BibitemOpen
  \bibfield  {author} {\bibinfo {author} {\bibfnamefont {E.~P.}\ \bibnamefont
  {Silaeva}}, \bibinfo {author} {\bibfnamefont {E.}~\bibnamefont {Bevillon}},
  \bibinfo {author} {\bibfnamefont {R.}~\bibnamefont {Stoian}}, \ and\ \bibinfo
  {author} {\bibfnamefont {J.~P.}\ \bibnamefont {Colombier}},\ }\href {\doibase
  10.1103/PhysRevB.98.094306} {\bibfield  {journal} {\bibinfo  {journal} {Phys.
  Rev. B}\ }\textbf {\bibinfo {volume} {98}},\ \bibinfo {pages} {094306}
  (\bibinfo {year} {2018})}\BibitemShut {NoStop}%
\bibitem [{\citenamefont {Weber}\ and\ \citenamefont
  {Rethfeld}(2017)}]{Weber2017}%
  \BibitemOpen
  \bibfield  {author} {\bibinfo {author} {\bibfnamefont {S.~T.}\ \bibnamefont
  {Weber}}\ and\ \bibinfo {author} {\bibfnamefont {B.}~\bibnamefont
  {Rethfeld}},\ }\href {\doibase 10.1016/j.apsusc.2017.02.183} {\bibfield
  {journal} {\bibinfo  {journal} {Applied Surface Science}\ }\textbf {\bibinfo
  {volume} {417}},\ \bibinfo {pages} {64} (\bibinfo {year} {2017})}\BibitemShut
  {NoStop}%
\bibitem [{\citenamefont {Waldecker}\ \emph {et~al.}(2016)\citenamefont
  {Waldecker}, \citenamefont {Bertoni}, \citenamefont {Ernstorfer},\ and\
  \citenamefont {Vorberger}}]{Waldecker2016}%
  \BibitemOpen
  \bibfield  {author} {\bibinfo {author} {\bibfnamefont {L.}~\bibnamefont
  {Waldecker}}, \bibinfo {author} {\bibfnamefont {R.}~\bibnamefont {Bertoni}},
  \bibinfo {author} {\bibfnamefont {R.}~\bibnamefont {Ernstorfer}}, \ and\
  \bibinfo {author} {\bibfnamefont {J.}~\bibnamefont {Vorberger}},\ }\href
  {\doibase 10.1103/PhysRevX.6.021003} {\bibfield  {journal} {\bibinfo
  {journal} {Phys. Rev. X}\ }\textbf {\bibinfo {volume} {6}},\ \bibinfo {pages}
  {021003} (\bibinfo {year} {2016})}\BibitemShut {NoStop}%
\bibitem [{\citenamefont {Ono}(2018)}]{Ono2018}%
  \BibitemOpen
  \bibfield  {author} {\bibinfo {author} {\bibfnamefont {S.}~\bibnamefont
  {Ono}},\ }\href {\doibase 10.1103/PhysRevB.97.054310} {\bibfield  {journal}
  {\bibinfo  {journal} {Phys. Rev. B}\ }\textbf {\bibinfo {volume} {97}},\
  \bibinfo {pages} {054310} (\bibinfo {year} {2018})}\BibitemShut {NoStop}%
\bibitem [{\citenamefont {Klett}\ and\ \citenamefont
  {Rethfeld}(2018)}]{Klett2018}%
  \BibitemOpen
  \bibfield  {author} {\bibinfo {author} {\bibfnamefont {I.}~\bibnamefont
  {Klett}}\ and\ \bibinfo {author} {\bibfnamefont {B.}~\bibnamefont
  {Rethfeld}},\ }\href {\doibase 10.1103/PhysRevB.98.144306} {\bibfield
  {journal} {\bibinfo  {journal} {Phys. Rev. B}\ }\textbf {\bibinfo {volume}
  {98}},\ \bibinfo {pages} {144306} (\bibinfo {year} {2018})}\BibitemShut
  {NoStop}%
\bibitem [{\citenamefont {Mueller}\ \emph {et~al.}(2013)\citenamefont
  {Mueller}, \citenamefont {Baral}, \citenamefont {Vollmar}, \citenamefont
  {Cinchetti}, \citenamefont {Aeschlimann}, \citenamefont {Schneider},\ and\
  \citenamefont {Rethfeld}}]{Mueller2013PRL}%
  \BibitemOpen
  \bibfield  {author} {\bibinfo {author} {\bibfnamefont {B.~Y.}\ \bibnamefont
  {Mueller}}, \bibinfo {author} {\bibfnamefont {A.}~\bibnamefont {Baral}},
  \bibinfo {author} {\bibfnamefont {S.}~\bibnamefont {Vollmar}}, \bibinfo
  {author} {\bibfnamefont {M.}~\bibnamefont {Cinchetti}}, \bibinfo {author}
  {\bibfnamefont {M.}~\bibnamefont {Aeschlimann}}, \bibinfo {author}
  {\bibfnamefont {H.~C.}\ \bibnamefont {Schneider}}, \ and\ \bibinfo {author}
  {\bibfnamefont {B.}~\bibnamefont {Rethfeld}},\ }\href {\doibase
  10.1103/PhysRevLett.111.167204} {\bibfield  {journal} {\bibinfo  {journal}
  {Phys. Rev. Lett.}\ }\textbf {\bibinfo {volume} {111}},\ \bibinfo {pages}
  {167204} (\bibinfo {year} {2013})}\BibitemShut {NoStop}%
\bibitem [{\citenamefont {Bierbrauer}\ \emph {et~al.}(2017)\citenamefont
  {Bierbrauer}, \citenamefont {Weber}, \citenamefont {Schummer}, \citenamefont
  {Barkowski}, \citenamefont {Mahro}, \citenamefont {Mathias}, \citenamefont
  {Schneider}, \citenamefont {Stadtm{\"u}ller}, \citenamefont {Aeschlimann},\
  and\ \citenamefont {Rethfeld}}]{Bierbrauer2017}%
  \BibitemOpen
  \bibfield  {author} {\bibinfo {author} {\bibfnamefont {U.}~\bibnamefont
  {Bierbrauer}}, \bibinfo {author} {\bibfnamefont {S.~T.}\ \bibnamefont
  {Weber}}, \bibinfo {author} {\bibfnamefont {D.}~\bibnamefont {Schummer}},
  \bibinfo {author} {\bibfnamefont {M.}~\bibnamefont {Barkowski}}, \bibinfo
  {author} {\bibfnamefont {A.-K.}\ \bibnamefont {Mahro}}, \bibinfo {author}
  {\bibfnamefont {S.}~\bibnamefont {Mathias}}, \bibinfo {author} {\bibfnamefont
  {H.~C.}\ \bibnamefont {Schneider}}, \bibinfo {author} {\bibfnamefont
  {B.}~\bibnamefont {Stadtm{\"u}ller}}, \bibinfo {author} {\bibfnamefont
  {M.}~\bibnamefont {Aeschlimann}}, \ and\ \bibinfo {author} {\bibfnamefont
  {B.}~\bibnamefont {Rethfeld}},\ }\href {\doibase 10.1088/1361-648X/aa6f73}
  {\bibfield  {journal} {\bibinfo  {journal} {Journal of Physics: Condensed
  Matter}\ }\textbf {\bibinfo {volume} {29}},\ \bibinfo {pages} {244002}
  (\bibinfo {year} {2017})}\BibitemShut {NoStop}%
\bibitem [{\citenamefont {Mueller}\ and\ \citenamefont
  {Rethfeld}(2014)}]{Mueller2014}%
  \BibitemOpen
  \bibfield  {author} {\bibinfo {author} {\bibfnamefont {B.~Y.}\ \bibnamefont
  {Mueller}}\ and\ \bibinfo {author} {\bibfnamefont {B.}~\bibnamefont
  {Rethfeld}},\ }\href {\doibase 10.1016/j.apsusc.2013.12.074} {\bibfield
  {journal} {\bibinfo  {journal} {Applied Surface Science}\ }\textbf {\bibinfo
  {volume} {302}},\ \bibinfo {pages} {24} (\bibinfo {year} {2014})}\BibitemShut
  {NoStop}%
\bibitem [{\citenamefont {Lide}\ \emph {et~al.}(2005)\citenamefont {Lide},
  \citenamefont {Baysinger}, \citenamefont {Berger}, \citenamefont {Goldberg},
  \citenamefont {Kehiaian}, \citenamefont {Kuchitsu}, \citenamefont
  {Rosenblatt}, \citenamefont {Roth},\ and\ \citenamefont
  {Zwillinger}}]{CRC05}%
  \BibitemOpen
  \bibfield  {author} {\bibinfo {author} {\bibfnamefont {D.~R.}\ \bibnamefont
  {Lide}}, \bibinfo {author} {\bibfnamefont {G.}~\bibnamefont {Baysinger}},
  \bibinfo {author} {\bibfnamefont {L.~I.}\ \bibnamefont {Berger}}, \bibinfo
  {author} {\bibfnamefont {R.~N.}\ \bibnamefont {Goldberg}}, \bibinfo {author}
  {\bibfnamefont {H.~V.}\ \bibnamefont {Kehiaian}}, \bibinfo {author}
  {\bibfnamefont {K.}~\bibnamefont {Kuchitsu}}, \bibinfo {author}
  {\bibfnamefont {G.}~\bibnamefont {Rosenblatt}}, \bibinfo {author}
  {\bibfnamefont {D.~L.}\ \bibnamefont {Roth}}, \ and\ \bibinfo {author}
  {\bibfnamefont {D.}~\bibnamefont {Zwillinger}},\ }\href@noop {} {\emph
  {\bibinfo {title} {{CRC Handbook of Chemistry and Physics}}}}\ (\bibinfo
  {publisher} {CRC Press, Boca Raton,FL},\ \bibinfo {year} {2005})\BibitemShut
  {NoStop}%
\bibitem [{\citenamefont {Czycholl}(2008)}]{Czycholl}%
  \BibitemOpen
  \bibfield  {author} {\bibinfo {author} {\bibfnamefont {G.}~\bibnamefont
  {Czycholl}},\ }\href@noop {} {\emph {\bibinfo {title} {Theoretische
  Festk\"{o}rperphysik}}},\ \bibinfo {edition} {3rd}\ ed.\ (\bibinfo
  {publisher} {Springer},\ \bibinfo {address} {Berlin Heidelberg},\ \bibinfo
  {year} {2008})\BibitemShut {NoStop}%
\bibitem [{\citenamefont {Cho}\ \emph {et~al.}(2011)\citenamefont {Cho},
  \citenamefont {Engelhorn}, \citenamefont {Correa}, \citenamefont {Ogitsu},
  \citenamefont {Weber}, \citenamefont {Lee}, \citenamefont {Feng},
  \citenamefont {Ni}, \citenamefont {Ping}, \citenamefont {Nelson},
  \citenamefont {Prendergast}, \citenamefont {Lee}, \citenamefont {Falcone},\
  and\ \citenamefont {Heimann}}]{Cho2011}%
  \BibitemOpen
  \bibfield  {author} {\bibinfo {author} {\bibfnamefont {B.~I.}\ \bibnamefont
  {Cho}}, \bibinfo {author} {\bibfnamefont {K.}~\bibnamefont {Engelhorn}},
  \bibinfo {author} {\bibfnamefont {A.~A.}\ \bibnamefont {Correa}}, \bibinfo
  {author} {\bibfnamefont {T.}~\bibnamefont {Ogitsu}}, \bibinfo {author}
  {\bibfnamefont {C.~P.}\ \bibnamefont {Weber}}, \bibinfo {author}
  {\bibfnamefont {H.~J.}\ \bibnamefont {Lee}}, \bibinfo {author} {\bibfnamefont
  {J.}~\bibnamefont {Feng}}, \bibinfo {author} {\bibfnamefont {P.~A.}\
  \bibnamefont {Ni}}, \bibinfo {author} {\bibfnamefont {Y.}~\bibnamefont
  {Ping}}, \bibinfo {author} {\bibfnamefont {A.~J.}\ \bibnamefont {Nelson}},
  \bibinfo {author} {\bibfnamefont {D.}~\bibnamefont {Prendergast}}, \bibinfo
  {author} {\bibfnamefont {R.~W.}\ \bibnamefont {Lee}}, \bibinfo {author}
  {\bibfnamefont {R.~W.}\ \bibnamefont {Falcone}}, \ and\ \bibinfo {author}
  {\bibfnamefont {P.~A.}\ \bibnamefont {Heimann}},\ }\href {\doibase
  10.1103/PhysRevLett.106.167601} {\bibfield  {journal} {\bibinfo  {journal}
  {Phys. Rev. Lett.}\ }\textbf {\bibinfo {volume} {106}},\ \bibinfo {pages}
  {167601} (\bibinfo {year} {2011})}\BibitemShut {NoStop}%
\bibitem [{\citenamefont {Anisimov}\ \emph {et~al.}(1974)\citenamefont
  {Anisimov}, \citenamefont {Kapeliovich},\ and\ \citenamefont
  {Perel'man}}]{Anisimov1974}%
  \BibitemOpen
  \bibfield  {author} {\bibinfo {author} {\bibfnamefont {S.~I.}\ \bibnamefont
  {Anisimov}}, \bibinfo {author} {\bibfnamefont {B.~L.}\ \bibnamefont
  {Kapeliovich}}, \ and\ \bibinfo {author} {\bibfnamefont {T.~L.}\ \bibnamefont
  {Perel'man}},\ }\href
  {http://www.jetp.ac.ru/cgi-bin/e/index/e/39/2/p375?a=list} {\bibfield
  {journal} {\bibinfo  {journal} {Sov. Phys. JETP}\ }\textbf {\bibinfo {volume}
  {39}},\ \bibinfo {pages} {375} (\bibinfo {year} {1974})}\BibitemShut
  {NoStop}%
\end{thebibliography}%
	
\end{document}